\documentclass[10pt,superscriptaddress,twocolumn,showkeys,aps,prb]{revtex4-1}
\usepackage{cancel}
\usepackage{mathrsfs}
\usepackage{bbold}
\usepackage{float}
\usepackage{graphicx}
\usepackage{bm}
\usepackage{amsmath} 
\usepackage{booktabs}

\def\nbOne{\pmb{\mathbb 1}}

\def\nbOne{{\mathbb 1}}

\def\be{\begin{equation}}
\def\ee{\end{equation}}
\def\beq{\begin{eqnarray}}
\def\eeq{\end{eqnarray}}

\RequirePackage{color}
\definecolor{MyDarkGreen}{rgb}{0.02,0.60,0.06}

\usepackage{ulem}

\begin{document}

\date{\today}

\title{Charge and spin polarized currents in mesoscopic rings with Rashba spin-orbit interactions coupled to an electron reservoir}
\author{M. Ellner}
\affiliation{Escuela de F\'isica, Facultad de Ciencias, Universidad Central de Venezuela, 1040 Caracas, Venezuela.}
\author{N. Bol\'ivar}
\affiliation{Escuela de F\'isica, Facultad de Ciencias, Universidad Central de Venezuela, 1040 Caracas, Venezuela.}
\affiliation{Groupe de Physique Statistique, Institut Jean Lamour, Universit\'e de Lorraine, 54506 Vandoeuvre-les-Nancy Cedex, France.}
\affiliation{Centro de F\'isica, Instituto Venezolano de Investigaciones Cient\'ificas, 21827, Caracas, 1020 A, Venezuela.}
\author{B. Berche}
\affiliation{Groupe de Physique Statistique, Institut Jean Lamour, Universit\'e de Lorraine, 54506 Vandoeuvre-les-Nancy Cedex, France.}
\affiliation{Centro de F\'isica, Instituto Venezolano de Investigaciones Cient\'ificas, 21827, Caracas, 1020 A, Venezuela.}
\author{E. Medina}
\affiliation{Centro de F\'isica, Instituto Venezolano de Investigaciones Cient\'ificas, 21827, Caracas, 1020 A, Venezuela.}
\affiliation{Groupe de Physique Statistique, Institut Jean Lamour, Universit\'e de Lorraine, 54506 Vandoeuvre-les-Nancy Cedex, France.}
\affiliation{Escuela de F\'isica, Facultad de Ciencias, Universidad Central de Venezuela, 1040 Caracas, Venezuela.}

\begin{abstract}
The electronic states of a mesoscopic ring are assessed in the presence of Rashba Spin Orbit coupling and a $U(1)$ gauge field. Spin symmetric coupling to an ideal lead is implemented following B\"uttiker's voltage probe. The exact density of states is derived using the reservoir uncoupled eigenstates as basis functions mixed by the reservoir coupling. The decay time of uncoupled electron eigenstates is derived by fitting the broadening profiles. The spin and charge persistent currents are computed in the presence of the SO interaction and the reservoir coupling for two distinct scenarios of the electron filling fraction. The degradation of the persistent currents depends uniformly on the reservoir coupling but nonuniformly in temperature, the latter due to the fact that currents emerge from different depths of the Fermi sea, and thus for some regimes of flux, they are provided with a protective gap. Such flux regimes can be tailored by the SO coupling for both charge and spin currents. 
\end{abstract}


\maketitle

\section{Introduction}
Recently there has been a growing interest in the Spin Orbit (SO) interaction, partly due to its omnipresence in non-centrosymmetric semiconductors with high technological value such as GaAs, InSb and CdTe, all with a Zinc-Blende structure\cite{Winkler}. It is of special interest that the Rashba Spin Orbit Interaction (RSOI) may be used to implement control of the spin degree of freedom through electrical means\cite{SpinControlElectrical}, since spin more weakly couples to decoherence effects as compared to the charge\cite{SpinDecoherence}. In particular, spin-asymmetric mesoscopic rings, combine well known charge interference effects with spin-orbit interactions, that cause spin splitting and spin interference\cite{Nitta} even in the absence of a magnetic field, while preserving time reversal symmetry. Such combination of interactions plus the existence of edges, give rise to the spin Quantum Hall Effect and topological insulators\cite{Kane}. These novel states of matter have many new potential applications radiating from the fact that conduction states are protected against impurity scattering.

Recent proposals, based on spin-orbit controlled spin precession in mesoscopics rings or interferometric devices, cover many mechanisms for generating spin polarized electrons by electric and magnetic flux control\cite{Governale,Frustaglia,Ionicio,Hatano,Duan} and charge and spin currents driven by electro-magnetic pulses\cite{Gudmundsson}. Graphene based materials for rings are also promising, due to the possibility of substrate interactions\cite{Dedkov} or intercalating atoms\cite{Marchenko} that have been devised to enhance an otherwise weak Rashba spin-orbit coupling.
 
In this work we study the effects of voltage probe coupling and temperature effects on the coherence of spin split
bands in a Rashba coupled ring. The experimental realization of the ring is generally understood to be within a two dimensional electron gas,
where the Rashba coupling is induced by structural inversion assymetry by a gate voltage. Nevertheless, this same gate can be a source
of dephasing as electrons couple to it as a voltage probe. The robustness of any proposed device must measure up to
the effects of the environment. A particularly simple model, for analytical treatment, is the B\"uttiker probe model\cite{Buttiker}, extensively
used in the literature\cite{Pareek,Tsymbal,LiYan,Golizadeh}. An emblematic phase coherent phenomenon used as a testing ground
is that of persistent charge and spin currents, the latter made possible by the RSO coupling. We determine the persistent charge and spin currents in a 2D electron gas built into a mesoscopic ring with narrow confinement\cite{Frustaglia,Sheng}. RSO interaction is contemplated as arising from structural inversion asymmetry built into the electron potential controlled by a gate. The solution to this problem in the completely coherent limit, has been addressed before both in the continuum Hamiltonian\cite{Nitta,Frustaglia,Sheng,Margulis,Duan} and the tight binding version\cite{RingTB}. We briefly revisit the problem in the continuum to generate the basis functions in order to address the exact solution to the voltage probe model\cite{Buttiker} including SO active media. While the uncoupled ring is diagonalisable as a Hamiltonian problem, the reservoir coupling can
be formulated in the scattering formalism. The coupling of these two problems, generalising B\"uttiker's treatment, allows us to obtain analytical expressions for the densities of states and equilibrium currents from the quantum mechanical definitions.

We compute the decay of persistent currents with the coupling to the electron reservoir and also with temperature, determined solely by
effect of the Fermi distribution. For low enough temperatures we find that charge and spin persistent currents exhibit robust oscillations
following the uncoupled spectrum of the ring and their magnitude can be controlled by the external magnetic flux (up to $~0.5 h/e$ through the ring).
The spin current can be made to switch signs and stay constant at constant magnitude quite robustly. While strong cancellation of the contributions to charge and spin currents are still generic in the presence of RSO coupling, we find that there are ranges in flux where currents are thermally protected by a gap. These ranges can be tuned by the SO coupling.

\section{States of the decoupled SO active ring}
The two dimensional quantum Hamiltonian for electrons of effective mass $m^*$ is given by
\be
H=\frac{{\bm\Pi}^2}{2m^*}+\alpha\left({\bm\sigma}\times{\bm \Pi}\right)+U({\bm r}),
\ee
were $\sigma_i$ are the Pauli matrices, ${\bm \Pi}=({\bm p}-e{\bm A})$ and $U(\bm r)$ defines the confinement potential of a ring geometry.
$\alpha$ is the coupling strength of the Rashba spin-orbit interaction ${\tilde V}_R$, tunable by an external electric field, and $A_i$ are the components of the vector potential associated with an external magnetic field in the $\hat z$ direction. It is assumed that only the ground state radial mode of the potential $U({\bm r})$ is involved. The treatment and role of higher order radial model has been treated in ref.~[\onlinecite{Sheng}].

\begin{figure}[ht]
\begin{center}
\includegraphics[scale=0.8]{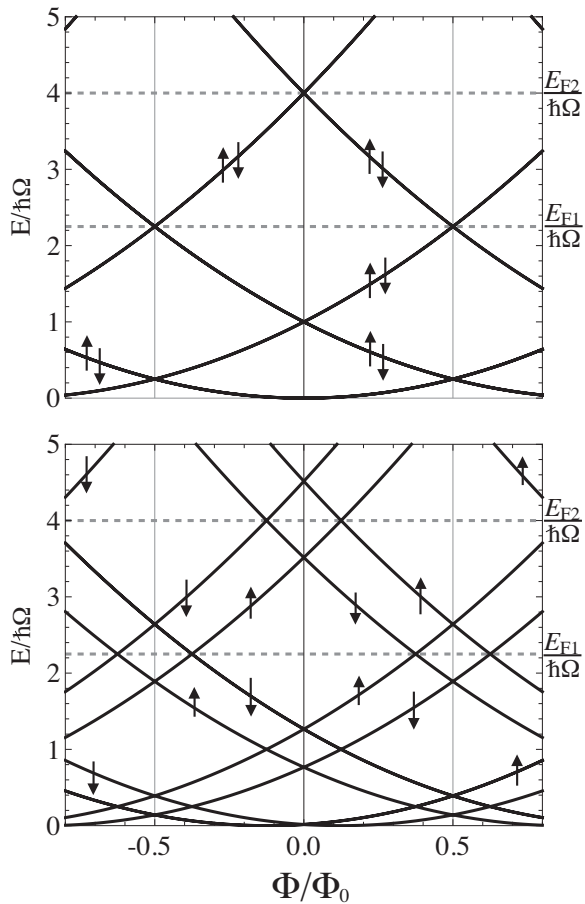}\\
\caption{\label{fig:DeS} Energy states of the decoupled ring for $\omega_{SO}/\Omega =0$ (top) and $\omega_{SO}/\Omega =0.75$ (bottom). The two dashed lines represent Fermi levels considered to compute the charge and spin currents below. The breaking of spin degeneracy, on applying SO coupling, allows for spin equilibrium currents.}
\end{center}
\end{figure}

A straightforward ``classical" coordinate change of this Hamiltonian, $(x,y)\rightarrow (\rho,\phi)$ results in a non-hermitian form that must be symmetrized appropriately. The correct hermitian RSO potential in polar coordinates is given by the usual coordinate transformation plus a basis rotation of the spinor \cite{Berch},
\be
V_{R}=e^{i\sigma_z \frac{\varphi}{2}}\tilde{V}_{R}e^{-i\sigma_z \frac{\varphi}{2}} = -\hbar \omega_{SO}\sigma_{\rho}\left( i \partial_\varphi + \frac{\Phi}{\Phi_0}\right)-i \hbar \frac{\omega_{SO}}{2}\sigma_\varphi,
\ee
where $\omega_{SO}=\frac{\alpha}{a}$, $a$ is the ring radius and $\Phi_0=2 \pi\hbar/e$ is the quantum of flux. The rotated Pauli matrices are defined as $\sigma_{\varphi}=-\sigma_x\sin\varphi+\sigma_y\cos\varphi$ and $\sigma_{\rho}=\sigma_x\cos\varphi+\sigma_y\sin\varphi$. Adding the kinetic energy operator reads the Hamiltonian
\be
H=\hbar\Omega\left(i\frac{\partial}{\partial\varphi}+\frac{\Phi}{\Phi_0}\right)^2-\hbar\omega_{SO}\sigma_{\rho}\left( i\frac{\partial}{\partial\varphi}+\frac{\Phi}{\Phi_0}\right)-i\frac{\hbar\omega_{SO}}{2}\sigma_{\varphi},
\ee
with $\Omega=\hbar/2ma^2$. Completing squares taking into account operator ordering and the angular dependencies of $\sigma_{\varphi}$ and $\sigma_\rho$, one arrives at the compact form,
\be
H=\hbar \Omega \left( -i \frac{\partial}{\partial \varphi}-\frac{\Phi}{\Phi_0}+ \frac{\omega_{SO}}{2\Omega}\sigma_\rho \right)^2 -\frac{\hbar\omega_{SO}^2}{4\Omega}.
\ee
In order to obtain the eigenvalues we can focus only on the quadratic term, and restore the additive scalar term to the resulting eigenvalue. We can then solve the simpler eigenvalue equation
\be
\left(-i\frac{\partial}{\partial\varphi}-\frac{\Phi}{\Phi_0}+\frac{\omega_{SO}}{2\Omega}\sigma_{\rho}\right)\psi=\sqrt{\frac{E}{\hbar\Omega}}\psi,
\ee
clearly $\psi$, a spinor, is also eigenfunction of the square of the previous operator with the square of the eigenvalue. The proposed form for the eigenspinor is
\be
\psi_j^{\mu}(\varphi)=e^{in_j^{\mu}\varphi}\chi^{\mu}(\varphi)=e^{in_j^{\mu}\varphi} \left(\begin{array}{ccc} A^{\mu}\\
e^{i\varphi }B^{\mu}
\end{array} \right),
\ee
where $j$ labels right and left propagating plane waves ($j=1$ clockwise and $j=2$ counterclockwise), $\mu$ is the spin label and $n_j^{\mu} \in \mathbb{Z}$ ($\mu=1$ spin up and $\mu=2$ spin down). Solving the matrix equation, the eigenvalues are found to be,
\be
E_{n,j}^{\mu}=\hbar \Omega\left((-1)^j n-\frac{\Phi}{\Phi_0}+\frac{1}{2\pi}\Phi_{AC}^{(\mu)}\right)^2 -\frac{\hbar\omega_{SO}^2}{4 \Omega} ,\label{ring-eigenvalues}
\ee
were $\Phi_{AC}=\pi\left(1+(-1)^{\mu}\sqrt{1+(\omega_{SO}/\Omega)^2}\right)$ (AC for Aharonov-Casher phase). The eigenfunction coefficients satisfy the relation
\be
\frac{\Omega}{\omega_{SO}}\left(1+(-1)^{\mu}\frac{1}{\cos\theta}\right) A^{\mu}=B^{\mu}, 
\ee
with $\cos\theta=1/\sqrt{1+(\omega_{SO}/\Omega)^2}$. One can then choose $A^{(1)}=B^{(2)}=\cos\frac{\theta}{2}$ and $-A^{(2)}=B^{(1)}=\sin\frac{\theta}{2}$. We thus arrive at the eigenfunctions
\begin{eqnarray}
\label{ring-eigenfunctions}
\psi _j^1(\varphi ) &=& {e^{in_j^1\varphi }}\left( {\begin{array}{*{20}{c}}
{\cos {\frac{\theta }{2}}}\\
{{e^{i\varphi }}\sin {\frac{\theta }{2}}}
\end{array}} \right),\nonumber \\
\psi _j^2(\varphi ) &=& {e^{in_j^2\varphi }}\left( {\begin{array}{*{20}{c}}
{\sin {\frac{\theta }{2}}}\\
{ - {e^{i\varphi }}\cos{\frac{\theta }{2}}}
\end{array}} \right), 
\end{eqnarray}
where $\frac{\theta}{2} ={{\tan }^{-1}}(\Omega/\omega_{SO}-\sqrt{\left(\Omega/\omega _{SO})^2+1\right)} )$. Figure~\ref{fig:DeS} shows the spectrum for $\omega_{SO}=0.75\Omega$.  The spin-orbit interaction alone preserves time reversal symmetry, so in the absence of a magnetic field $E_{n,+}^{\uparrow}=E_{n,-}^{\downarrow}$ i.e. twofold degeneracies. At half integer flux quanta this degeneracy is repeated. For other values of the flux the degeneracy is broken. For zero SO coupling and in the absence of a Zeeman term there is a peculiar twofold degeneracy for each level due to the closing of the wave function for half integer spin\cite{Bolivar}. Thus $E_{n,-}^{\uparrow}=E_{n+1,-}^{\downarrow}$ and $E_{n,+}^{\downarrow}=E_{n+1,+}^{\uparrow}$ for all fluxes. At zero and half integer flux quanta we have fourfold degeneracy in the absence of SO coupling. Such degeneracies are important when computing the corresponding charge and spin currents.

Wavefunctions in Eqs.~(\ref{ring-eigenfunctions}), form a complete four function basis to represent couplings of the system with an external voltage probe.

\section{Decoherence with Spin Orbit coupling}
In reference [\onlinecite{Buttiker}] B\"uttiker introduced an ingenious way to couple a simple quantum system (a ring) to reservoir that behaved like a voltage probe (zero current condition to reservoir). This model lead to a variety of generalizations\cite{reviewonprobes}. The approach here is similar; as the coupling to the reservoir is not defined in Hamiltonian terms and leads to dephasing, we have a Hamiltonian solution to the uncoupled problem and a scattering approach for the coupling to the reservoir. The two problems meet when using the complete basis of the uncoupled problem with the coefficients of these basis components determined by matching boundary conditions.

Coupling to the reservoir is introduced in a ring through an ideal lead that acts as a voltage probe (no net current threads the lead). The reservoir emits electrons with a Fermi distribution and absorbs electrons of any energy. Dephasing occurs due to the absence of a phase relation between injected and emitted electrons at a particular energy. 

 The coupling between the lead and the ring is described by the scattering matrix $S$ which relates the incoming and outgoing amplitudes $\vec{\alpha}'=S \vec{\alpha}$. The current conservation implies that $S$ is unitary, the matrix is $3 \times 3$ for each spin label $\mu$ as the coupling to the reservoir is spin symmetric. In general the matrix $S$ will dependent on five independent parameters, considering $S$ to be symmetric with respect to the two branches of the ring, the number of independent parameters reduces to three. 
\begin{equation}
\label{matrixS}S = \left( {\begin{array}{*{20}{c}}
r_{33}&t_{32}&t_{31}\\
t_{23}&r_{22}&t_{21}\\
t_{13}&t_{12}&r_{11}
\end{array}} \right)=\left( {\begin{array}{*{20}{c}}
{ - (a + b)}&{\sqrt \varepsilon  }&{\sqrt \varepsilon  }\\
{\sqrt \varepsilon  }&a&b\\
{\sqrt \varepsilon  }&b&a
\end{array}} \right),
\end{equation}
where $a=(\sqrt{1-2\varepsilon }-1)/2$ , $b=(\sqrt{1-2\varepsilon }+1)/2$ and  $\varepsilon$ is the coupling parameter with the reservoir, which varies  between $0$ and $1/2$ for the uncoupled and fully coupled limits respectively\cite{Buttiker2}. We have also written the $S$ matrix in terms of $t_{i,j}$, the transmission amplitude between the $i$-th and $j$-th lead and $r_{i,i}$ the reflection amplitude back into the same lead, with $i=1,2,3$, where $3$ refers to the reservoir lead and $1,2$ to the ring, either the left or the right to the reservoir lead.  The symmetry of the terms in the $S$ matrix, referred to this formulation depends on which fields are present, as we will see below.

The lead coupling the ring to the reservoir needs two equivalent spin channels and thus can be expanded as 
\begin{equation}
\label{spinorlead}\begin{array}{*{20}{c}}
{{\psi _{lead}}(x) = \sum\limits_{\mu  = 1,2} {{\phi _{lead}}} (x){\chi ^{(\mu )}}(0)}&{}&{x \in ( - \infty ,0]}
\end{array}.
\end{equation}
where $x$ is the coordinate along the lead and $x=0$ is defined as the coordinate at which the lead connects to the ring, while the reservoir is at $x=-\infty$, and $\chi^{\mu}$ is a two component spinor eigenstate of the $\sigma_z$. As the lead is not spin-orbit active the energies are $E=\hbar^2k^2/2m$. The coefficients of the expansion in Eq.~(\ref{spinorlead}) are given by, 
\be
\label{leadwavefunction}{\phi _{lead}}(x) = \sqrt {\mathcal{N}} \left( {{e^{ikx}} + {C_3}{e^{ - ikx}}} \right).
\ee
The normalization pre-factor is determined following B\"uttiker's argument: in an energy interval $E,E+dE$, the differential of current injected into the lead is $dI=ev(dN/dE)f(E)dE$, where $f(E)$ is the Fermi distribution, $dN/dE=1/2\pi\hbar v$ is the density of states of a perfect lead, and $v=\hbar k/m$. The wave function for the lead contemplates the correct current if $\mathcal{N}=f(E)dE/2\pi\hbar v$.

For the ring wave function, it is now a mixture of the four basis functions of the uncoupled case, so that we may accommodate for the new boundary conditions, we define
\be
\Psi(\varphi)=C_1^1\psi_1^1(\varphi)+C_1^2\psi_1^2(\varphi)+C_2^1\psi_2^1(\varphi)+C_2^2\psi_2^2(\varphi).
\ee
The coefficients are to be fixed by imposing equality of the wave functions at $x=0$ for $\varphi=0$ and $2\pi$.
The dispersion problem is written as $\vec{\alpha}'^{(\mu)}=S\vec{\alpha}^{(\mu)}$, the coefficients $\vec{\alpha}^{(\mu)}=(\alpha^{(\mu)},\beta^{(\mu)},\gamma^{(\mu)})$ and $\vec{\alpha}'^{(\mu)}=(\alpha'^{(\mu)},\beta'^{(\mu)},\gamma'^{(\mu)})$ are found evaluating (\ref{leadwavefunction}) at the junction at $x=0$ for $\alpha^{(\mu)}$ and $\alpha'^{(\mu)}$, and evaluating $\psi^\mu_2$ in $\varphi=0,2 \pi$ for the $\beta'^{(\mu)}$ and $\gamma^{(\mu)}$ respectively. The coefficients  $\beta^{(\mu)}$ and $\gamma'^{(\mu)}$ evaluating $\psi^\mu_1$ in $\varphi=0,2\pi$ respectively. The set of equations can be cast, for each spin subspace as
\begin{eqnarray}
\label{SSO}&&\left( {\begin{array}{*{20}{c}}
\sqrt{\mathcal N}C_3^\mu\\
C_1^{\mu}\\
C_2^{\mu}
\end{array}} \right)=\nonumber\\
 &&\left( {\begin{array}{*{20}{c}}
{ - (a + b)}&\sqrt{\varepsilon}&\sqrt{\varepsilon}e^{2\pi i n_1^\mu}\\
\sqrt{\varepsilon}&a&be^{2\pi i n_1^\mu}\\
\sqrt{\varepsilon}e^{-2\pi i n_2^\mu}&be^{-2\pi i n_2^\mu}&ae^{-2\pi i (n_1^{\mu}-n_2^{\mu})} 
\end{array}} \right)\left( {\begin{array}{*{20}{c}}
{\sqrt {\mathcal N} }\\
C_2^{\mu}\\
C_1^{\mu}
\end{array}} \right).\nonumber\\
\end{eqnarray}
where we have absorbed the phase factors into a redefined $S$ matrix that manifestly displays the symmetry of the system. Note that we can invert for the quantum number as a function of the energy and fields
\be
n_j^{\mu}=(-1)^j\sqrt{\frac{E}{\hbar\Omega}}+\frac{\Phi}{\Phi_0}-\frac{1}{2}\left(1+(-1)^{\mu}\sqrt{1+\left(\frac{\omega_{SO}}{\Omega}\right )^2}\right).
\ee
Referring to Eq.~(\ref{matrixS}) one can readily check that, in the absence of magnetic or SO fields, $t_{jk}=t_{kj}=\sqrt{\varepsilon}e^{2\pi i n_1^{\mu}}=\sqrt{\varepsilon}e^{-2\pi i n_2^{\mu}}$ i.e. $S$ is an orthogonal (symmetric) matrix, time reversal invariant. When the magnetic field is on but there is no SO coupling, then $t_{jk}\ne t_{kj}$ so $n_1^{\mu}\ne n_2^{\mu}$ and time reversal symmetry is broken. When the magnetic field is turned off and the SO coupling is present, time reversal symmetry is restored, and there is the additional symmetry for changing $j$ and $\mu$ labels simultaneously. Thus the larger $6\times6$ matrix $S\otimes\nbOne_s$ matrix is symplectic and embodies Kramers degeneracy.
Solving the system of equations one can obtain each of the amplitudes
\begin{eqnarray}
C_1^{\mu}&=&\frac{\sqrt{\epsilon\mathcal{N}}\left(1-e^{2\pi i n_2^{\mu}}\right)}{\left(1-be^{2\pi i n_1^{\mu}}\right)\left(b-be^{2\pi i n_2^{\mu}}\right)+a^2\left(1-be^{2\pi i n_1^{\mu}}\right)},\nonumber \\
C_2^{\mu}&=&\frac{\sqrt{\epsilon\mathcal{N}}\left(e^{2\pi i n_1^{\mu}}-1\right)}{\left(1-be^{2\pi i n_1^{\mu}}\right)\left(b-be^{2\pi i n_2^{\mu}}\right)+a^2\left(1-be^{2\pi i n_1^{\mu}}\right)},\nonumber \\
C_3^{\mu}&=&\frac{\epsilon \left(e^{2\pi i n_1^{\mu}}-1+\left(1-e^{2\pi i n_2^{\mu}}\right)e^{2\pi i n_1^{\mu}}\right)}{\left(1-be^{2\pi i n_1^{\mu}}\right)\left(b-be^{2\pi i n_2^{\mu}}\right)+a^2\left(1-be^{2\pi i n_1^{\mu}}\right)}\nonumber\\
&-& (a+b).
\end{eqnarray}
For the charge density the modulus squared of the coefficients acquire a particularly simple form in terms of the coupling parameters,
\begin{equation}
\label{modc1}|C_1^\mu {|^2} = \frac{{2\varepsilon {\mathcal N}}}{{{g^{(\mu )}}}}\left( {1 - \cos \left( {2\pi n_2^\mu } \right )} \right),
\end{equation}
\begin{equation}
\label{equation}|C_2^\mu {|^2} = \frac{{2\varepsilon {\mathcal N}}}{{{g^{(\mu )}}}}\left( {1 - \cos \left( {2\pi n_1^\mu } \right)} \right),
\end{equation}
\begin{equation}
\label{equation2}|C_3^\mu {|^2} = 1,
\end{equation}
where
\begin{widetext}
\begin{eqnarray}
{g^{(\mu )}} &=& 3 + \sqrt {1 - 2\varepsilon }  - 3\varepsilon  - 2\left( {1 + \sqrt {1 - 2\varepsilon }  - \varepsilon } \right)\cos \left( {2\pi n_1^\mu } \right) + 2\sqrt {1 - 2\varepsilon } \cos \left({2\pi \left( {n_1^\mu  - n_2^\mu } \right)} \right)+ \nonumber\\
 &-& 2\cos \left( {2\pi n_2^\mu } \right) + \cos \left( {2\pi \left( {n_1^\mu  + n_2^\mu } \right)} \right) + \left( {\sqrt {1 - 2\varepsilon }  - \varepsilon } \right)\left( { - 2\cos \left( {2\pi n_2^\mu } \right) + \cos \left( {2\pi \left( {n_1^\mu  + n_2^\mu } \right)} \right)} \right).
\end{eqnarray}
\end{widetext}
Note the very important character of the model expressed in Eq.~(\ref{equation2}); the lead amplitude has modulus one, thus two opposite propagating waves superpose to give a constant amplitude, which means there is no net current (voltage probe condition) to or from the reservoir.

For the density of states (DOS) we know that the number of electrons in the energy interval $dE$ is given by $dN = \left| {C_1^1} \right|^2 + \left| {C_1^2} \right|^2 + \left| {C_2^1} \right|^2 + \left| {C_2^2} \right|^2$. As each amplitude modulus is proportional to the energy interval $dE$ and using the chain rule $dN/dk=(dN/dE) (dE/dk)=(dN/dE) \hbar^2 k/m$. The number of electrons per unit energy range is given by
\be
\frac{dN}{dE}=\sum_{i,\mu}\frac{\varepsilon f(E)}{\pi \hbar v} \frac{(1-\cos 2\pi n_i^{\mu})}{g^{(\mu)}},
\ee
so the DOS can be written as
\begin{widetext}
\be
\label{DOD}\frac{dN}{dk}=\frac{2\varepsilon}{\pi}\left(\frac{\sin^2\left(2\pi n_1^1\right)+\sin^2\left(2\pi n_2^1\right)}{g^{(1)}} 
 + \frac{\sin^2\left(2\pi n_1^2\right)+\sin^2\left(2\pi n_2^2 \right)}{g^{(2)}} \right).
\ee
\end{widetext}
The explicit relation between DOS and energy comes from substituting the expressions for $n_j^{\mu}=(-1)^j\sqrt{E/\hbar\Omega}+\Phi/\Phi_0-1/2\left(1+(-1)^{\mu}\sqrt{1+(\omega_{SO}/\Omega)^2}\right)$ from the uncoupled problem. These expressions now define this quantum number which becomes a continuous function of the energy and flux and SO coupling, no longer restricted to be integer or half integer, as the problem is coupled.

\begin{figure}[ht]
\begin{center}
\includegraphics[scale=0.9]{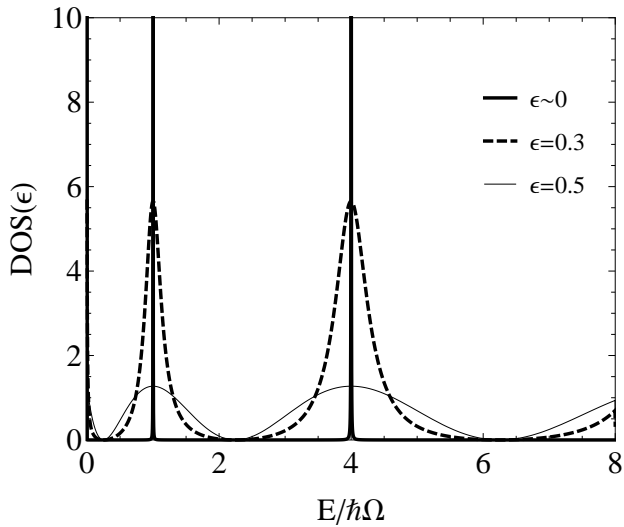}\\
\caption{\label{fig:wf} Density of states on the ring as a function of the energy for two values of the coupling parameter of the reservoir and $T=0$. (${{\omega }_{SO}}/\Omega =0$, $\Phi /{{\Phi }_{0}}=0$). The energy is expressed in units ${{E}_{0}}=\hbar \Omega$.}
\end{center}
\end{figure}

\begin{figure}[ht]
\begin{center}
\includegraphics[scale=0.9]{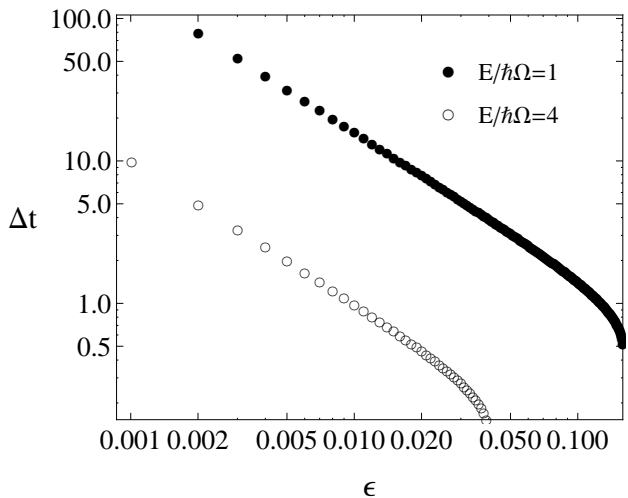}\\
\caption{\label{fig:pl} Lifetime of the electrons in the ring as a function of the coupling to the reservoir $\varepsilon$. The energies correspond to the quantized values of the decoupled states. The time is given in atomic units $1~{\rm a.u.} \approx 2.4 \times 10^{-17}{\rm s} $.}
\end{center}
\end{figure}
\begin{figure}[ht]
\begin{tabular}{cc}
a)\includegraphics[scale=0.85]{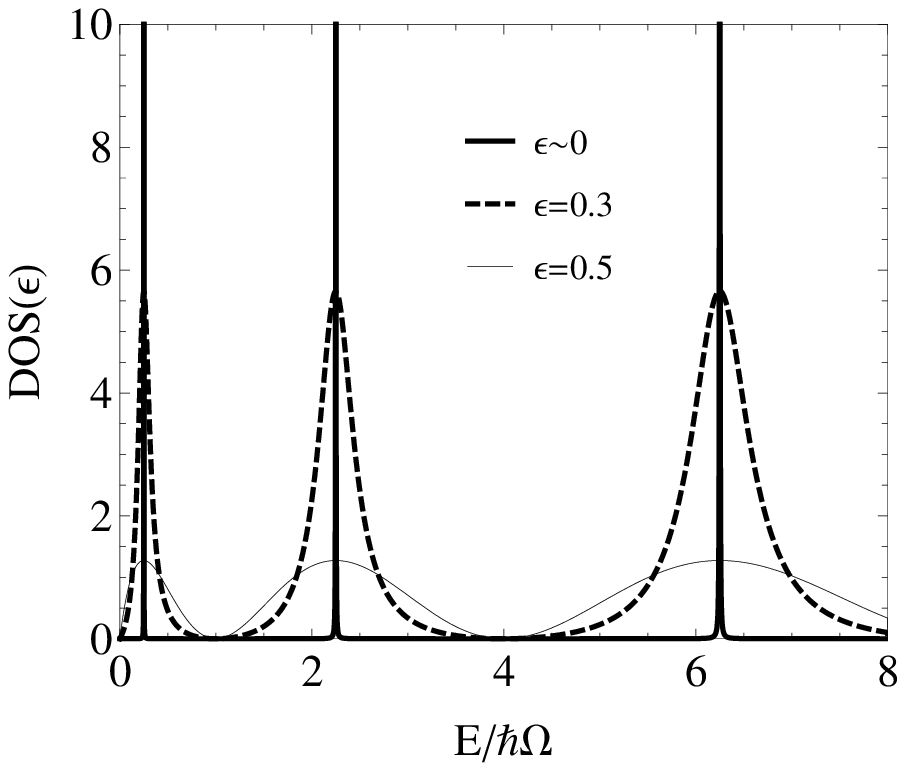}\\
\\
b)\includegraphics[scale=0.85]{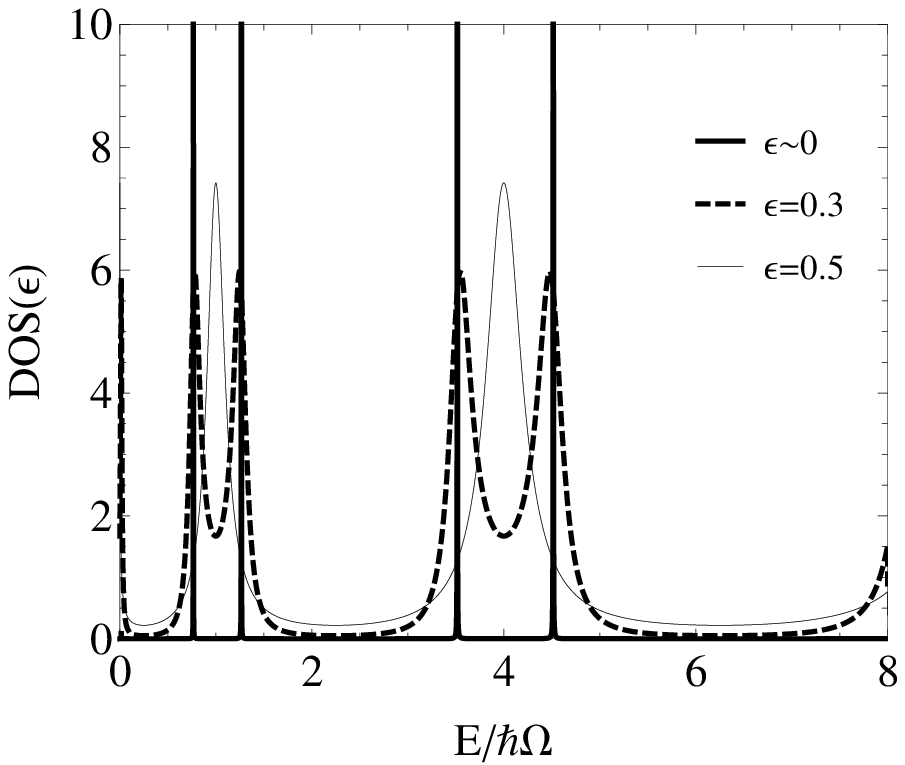}
\end{tabular}
\caption{\label{fig:DDE2} Density of states of the ring as a function of the energy for two values of $\varepsilon$ and $T=0$. The value of the parameters are a) ${{\omega }_{SO}}/\Omega =0$, $\Phi /{{\Phi }_{0}}=0.5$ and b) ${{\omega }_{SO}}/\Omega =0.75$ , $\Phi /{{\Phi }_{0}}=0$.}
\end{figure}

The limit of zero fields (neither SO nor magnetic field) with coupling to the reservoir recovers B\"uttiker's result\cite{Buttiker},
\[\frac{{dN}}{{dk}} =  \frac{{4\varepsilon \cos^2 \left({2\pi \sqrt {\frac{E}{{\hbar \Omega }}} } \right) }}{{\pi \left( { - 1 + \varepsilon  + \sqrt {1 - 2\varepsilon } \cos \left({2\pi \sqrt {\frac{E}{{\hbar \Omega }}} } \right)} \right)}}.\] 
Fig. \ref{fig:wf} show the DOS for $\varepsilon \neq 0$.  The levels increasingly broaden around the quantized energies of the decoupled ring ($\varepsilon = 0$) as $\varepsilon$ increases. The uncoupled quantized values correspond to the poles of the density of states at zero coupling, which obey the relation $E=m^2\hbar \Omega$, with $m$ an integer (values $m^2=0,1,4..$ in figure). When the coupling is turned on, the levels are shifted to lower energies as they broaden, as expected in general for complex self energy corrections\cite{Pastawski}.
Deeper levels are less coupled to the reservoir than the shallower counterparts since there is partial transmission to the reservoir lead. 

Making a correspondence between level broadening and electron lifetime by fitting the resonance to a Lorentzian form leads to Figure~\ref{fig:pl}. Only the regime where the broadening is reasonably Lorentzian is taken into account. As can be seen from the figure, the broadening function becomes non-trivial for $\varepsilon>0.1$, where the power-law decay changes. A power-law decay of the lifetime with the reservoir coupling is observed only for the smaller couplings. In spite of the large smearing of the energy levels and the exact treatment, this model reservoir always
yields a wavefunction. 

The magnetic field shifts the states and the SO coupling breaks the twofold degeneracy as was discussed.  Figs.~\ref{fig:DDE2}a and  \ref{fig:DDE2}b, show the effect of the magnetic field and the SO coupling, respectively, on the DOS. In panel a) each peak is doubly degenerate, while this degeneracy is broken with SO as depicted in panel b). The values assumed for the RSO e.g. $\omega_{SO}/\Omega=0.75$ which corresponds to $\hbar\alpha \sim 3.02\times10^{-12}$ eV.m, a realistic value for GaAs\cite{Datta}.  This degeneracy can appear to exist when the coupling to the reservoir is sufficiently large (see panel b) for $\varepsilon=0.5$, as the DOS broadens into a single peak containing both levels.

\section{Persistent charge currents}
For a decoupled ring at zero temperature, the charge persistent currents can be calculated by the linear response relation \cite{Sangchul,Berch}
$J_q=-\sum_i \frac{dE_i}{d \Phi}$ where $i$ encompasses the occupied states. The leading contribution to the current, due to cancellation of current contributions from state with opposite slopes, are the states close to the Fermi level. The linear response relation is not useful for the case we have coupling to the reservoir, since the energy broaden into a continuum of levels. On the other hand we have derived the exact wave functions from which the current may be determined by the expectation value of the charge current operator $\Psi^{\dag}ev_{\varphi}\Psi$ where
\begin{eqnarray}
\label{velocityop}
v_{\varphi}&=& a\dot{\varphi}= (a/i\hbar) [\varphi,H]\nonumber\\
&=&-2a\Omega\left(-i\frac{\partial}{\partial\varphi}-\frac{\Phi}{\Phi_0}+\frac{\omega_{SO}}{2\Omega}\sigma_{\rho}\right),
\end{eqnarray} 
and integrating over all occupied states up to the Fermi level including the electron occupation numbers.
\begin{eqnarray}
J_q=-\frac{2\varepsilon\hbar\Omega}{\Phi_0}\sum_{m,\mu}\int \frac{dE}{\hbar\Omega}\frac{f(E)}{\sqrt{\frac{E}{\hbar\Omega}}~g^{\mu}}&&\sin^2(\pi n_m^{\mu})\nonumber\\
\times&&\left[n_{\overline{m}}^{\mu}-\frac{\Phi}{\Phi_0}+\delta^{\mu}\right],\nonumber\\
\end{eqnarray}
with
\begin{equation}
\delta^{1}=\sin^2\frac{\theta}{2}+\frac{\omega_{SO}}{2\Omega}\sin\theta;~~~~\delta^{2}=\cos^2\frac{\theta}{2}-\frac{\omega_{SO}}{2\Omega}\sin\theta,
\end{equation}
\begin{figure}[ht]
\centering
\begin{tabular}{cc}
\includegraphics[scale=0.92]{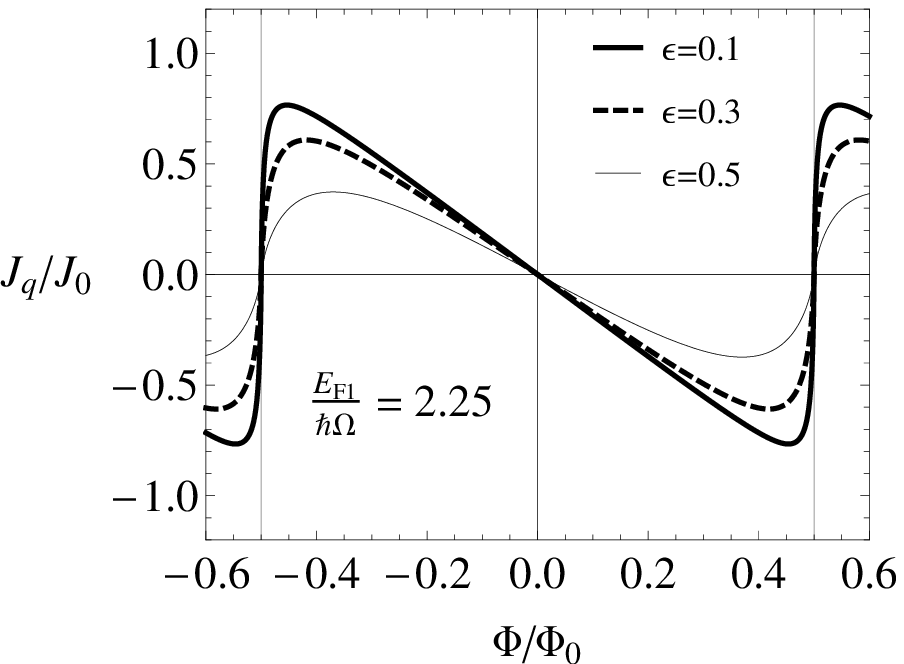} \\
\\
\includegraphics[scale=0.92]{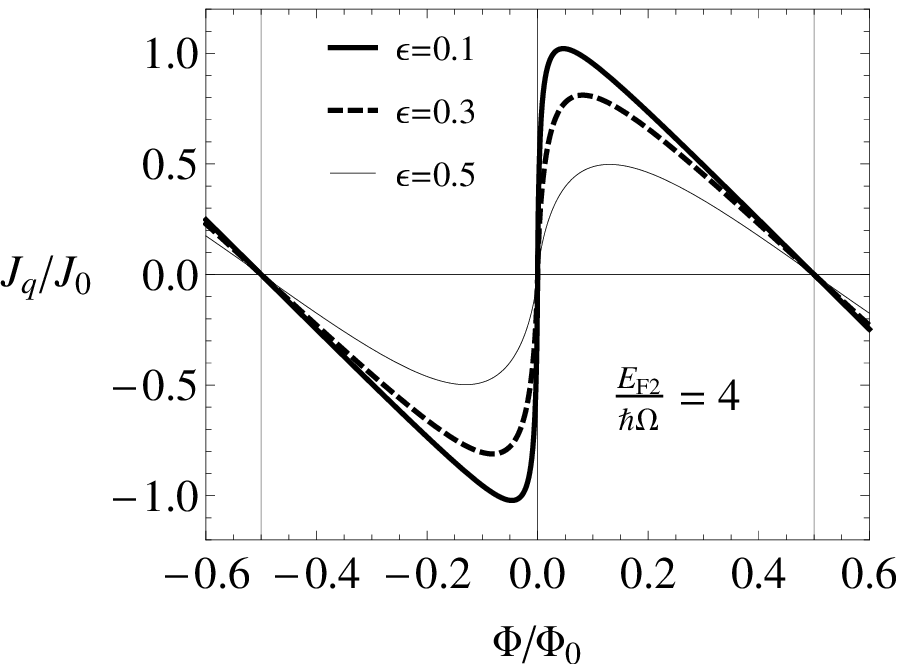}
\end{tabular}
\caption{\label{fig:PCC} Charge persistent current as a function of the magnetic flux for three values of the reservoir coupling parameter. The number of electrons is $6$ (top) and $8$ (bottom) corresponding to a Fermi energy of $E_{F_1}$ and $E_{F_2}$ respectively (see Fig.\ref{fig:DeS}). The RSO interaction is off and the persistent current is given in units $J_{0}=\hbar\Omega/\Phi_0$.}
\end{figure}
where $\overline{m}$ is the complement value of $m$ and a natural current scale $J_0=\hbar\Omega/\Phi_0$ is identified. Note that $\varepsilon=0$ does not imply zero current\cite{Buttiker} (in fact it is largest at zero coupling) as $g^{\mu}$ also depends on the coupling with a nontrivial limit behaviour. We will separate the discussion into two cases: I) The Fermi level fixes $N=6$ electrons, (see Fig \ref{fig:PCC} panel a)) and II) $N=8$, (see Fig \ref{fig:PCC} panel b)).  In the absence of RSO interaction for the first case, there are two electrons, one with spin up and the other with spin down at each energy. At the Fermi level, two bands which describe electrons with different propagation numbers $j$ cross each other at half integer steps in $\Phi_0$ (see Fig.\ref{fig:DeS}). This results in a jump in the sign of the current at these values. In the second case the levels cross at zero or integer flux quanta, and the sign jump occurs at those points. These are the behaviours expected also for small couplings to the reservoir. Fig.\ref{fig:PCC} shows the charge currents without the SO coupling as a function of the magnetic field. The reduction in amplitude of the current as a function of the coupling strength is evident as decoherence increases. For Fermi level $E_{f_1}$ the persistent current is minimal for the smallest fluxes and gradually grows, while for $E_{f_2}$ the current is maximal at the smallest fluxes and decreases thereof.

\begin{figure}[ht]
\begin{tabular}{cc}
\includegraphics[scale=0.91]{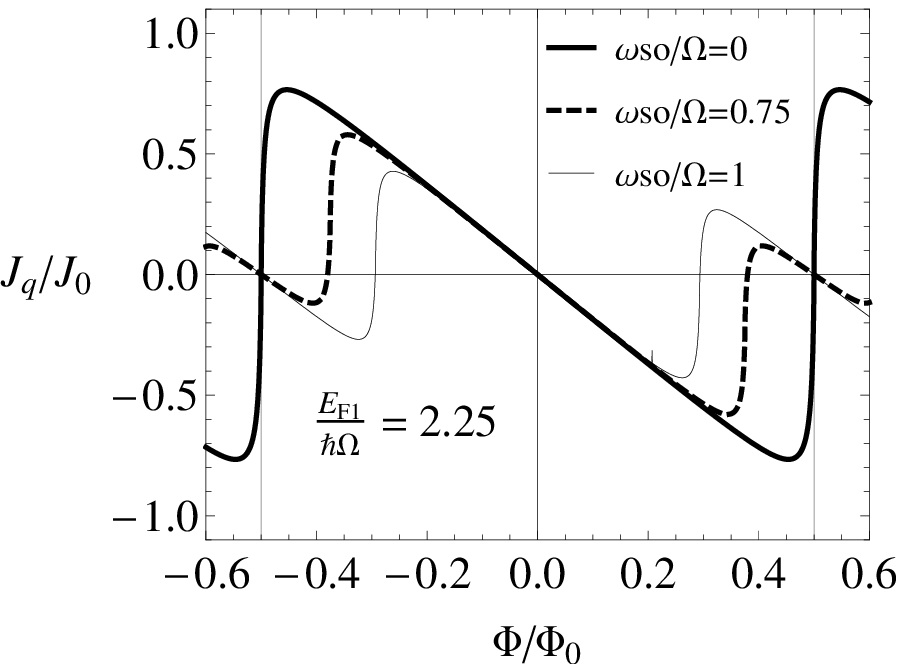} \\
\\
\includegraphics[scale=0.91]{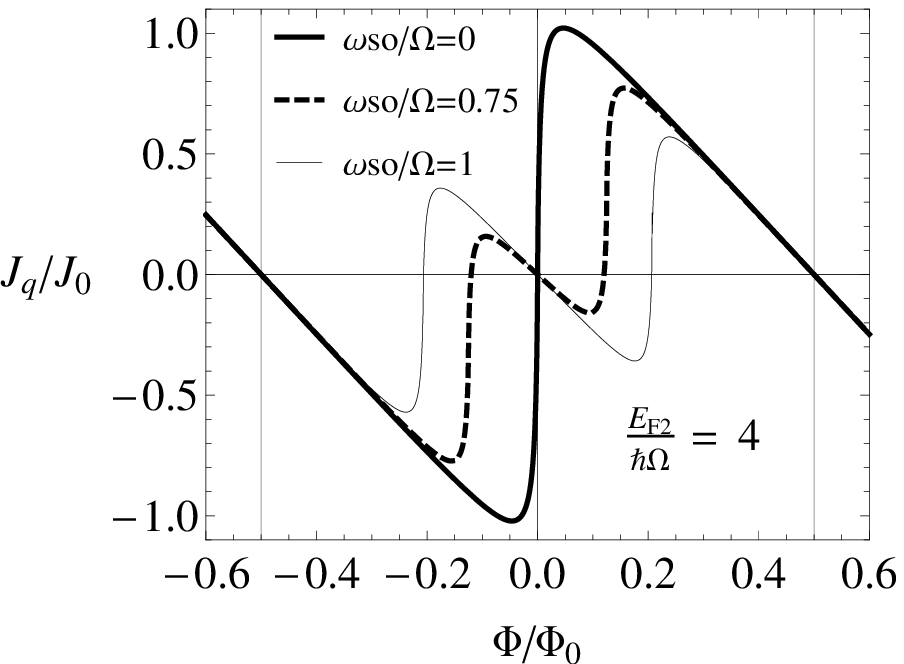}
\end{tabular}
\caption{\label{fig:PCCSO} Charge persistent current as a function of the magnetic flux for different RSO values. The reservoir coupling is $\varepsilon  = 0.1$. The number of electrons is $6$ (top) and $8$ (bottom) corresponding to a Fermi energy of $E_{F_1}$ and $E_{F_2}$ according to Fig.\ref{fig:DeS}}
\end{figure}

After including RSO, the crossing between bands at the Fermi level shift to $\Phi/\Phi_0=m/2+(1 \pm \sqrt{1+(\omega_{SO}/\Omega)^2})/2$, with $m \in \mathbb{Z}$ for the case I) and  $\Phi/\Phi_0=m/2\pm \sqrt{1+(\omega_{SO}/\Omega)^2}/2$ for case II) displacing the current jumps and introducing two more for each of the Fermi level scenarios\cite{Ekenberg} (see Fig \ref{fig:PCCSO}).  The current jumps from $|\Phi/\Phi_0|<0.5$ have a smaller amplitude at finite RSO, because the levels in the latter case are non degenerate, and cause only half of the full
current jump amplitude at $|\Phi/\Phi_0|=0.5$.

\begin{figure}[ht]
\begin{tabular}{cc}
\includegraphics[scale=0.85]{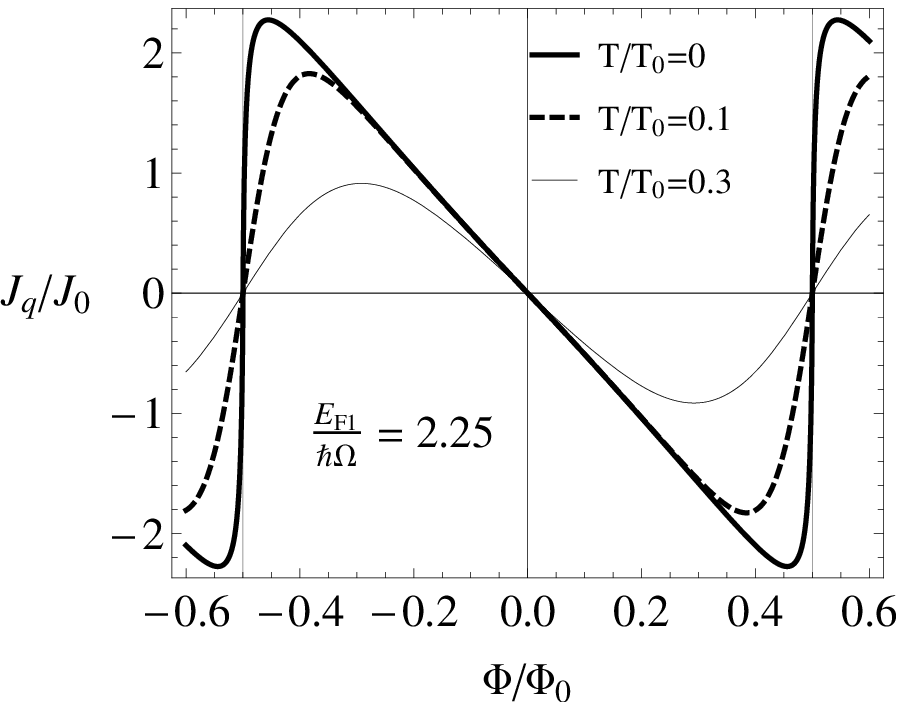} \\
\\
\includegraphics[scale=0.85]{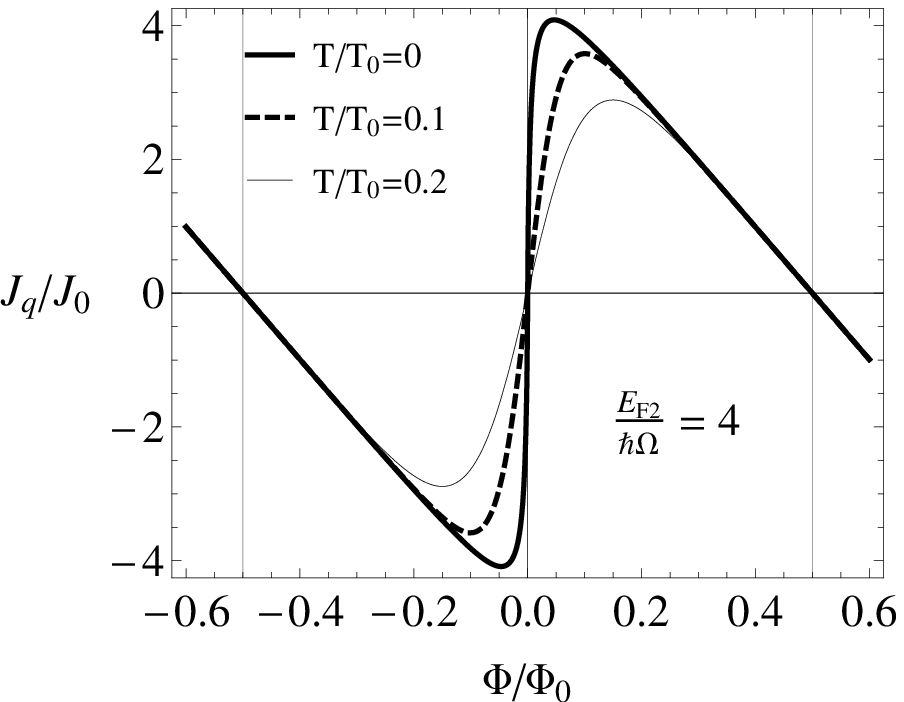}
\end{tabular}
\caption{\label{fig:CTEMP} Charge persistent current as a function of the magnetic flux, for different temperatures, for a fixed $\varepsilon  = 0.1$ and referred to the scale $T_0=\hbar\Omega/k_B$. The number of electrons is $6$ (top) and $8$ (bottom). Note the low sensitivity of the current to thermal effects when current arises form below the Fermi levels.}
\end{figure}

\begin{figure}[ht]
\includegraphics[scale=0.85]{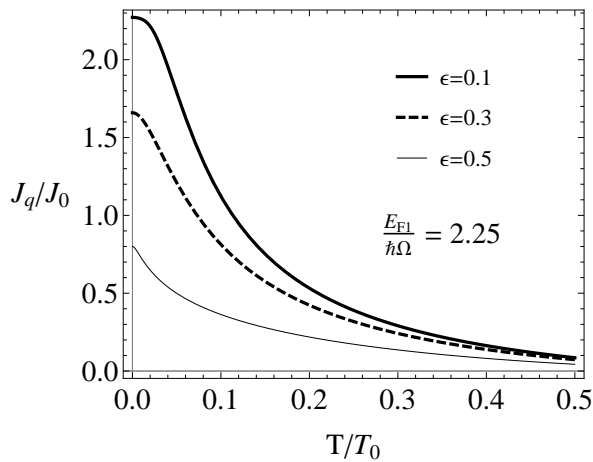}
\caption{\label{fig:CuVsT2} Temperature dependence of the charge current for the scenario of $E_{F_1}$. Depending on the magnetic flux chosen, the current can be degraded completely.}
\end{figure}

The degradation of current with temperature has a distinctive character as compared to the coupling to the reservoir, as can be seen in Fig.\ref{fig:CTEMP}. The temperature effect will be small when the current emanates from a level appreciably below the Fermi level, so that few electrons are actually promoted to counter current states. On the other hand, for fluxes where the currents arise from levels close to the Fermi level, the currents quickly degrade. For the case where currents originate from within the Fermi sea,  there is a gap protecting persistent currents that is energy dispersion dependent. See Refs. \onlinecite{Bolivar} and \onlinecite{Daniel} where currents are protected from thermal effects by the linear dispersion. Figure~\ref{fig:CuVsT2} shows the dependence of charge current on temperature, for different ring-reservoir couplings. For certain ranges of the magnetic flux, the persistent current can be degraded completely. We estimate the magnitude of the thermal effects by using the temperature scale $T_0=\hbar\Omega/k_B$, As $\Omega$ depends on the size of the ring, $T/T_0=0.5$ in the Figures, correspond to temperatures between 526 mK and 59 mK for ring sizes between 100 nm and 300 nm and an effective mass of $m^*=0.042 m_e$. This implies that the gap for persistent current degradation is of the order of 40~$\mu$eV for the smallest of the rings. Improving this gap with either effective mass of ring radius and flux point of operation, might improve the thermal robustness of high sensitivity cantilevers\cite{Shanks} for noise and electron thermometry.

\section{Persistent Spin currents}
The standard calculation is through the anticommutator of the velocity with the spin operator\cite{Molnar,Berch,Sangchul},
\[{J_s^{z}} = \frac{\hbar }{4}\Psi^\dag\{\sigma_z, v_{\varphi}\}\Psi.\]
Invoking the full wave function derived above for the coupled ring and the velocity operator in Eq.~(\ref{velocityop}) one can derive the spin current as
\begin{eqnarray}
J_s^{z}=-\varepsilon\hbar\Omega \sum_{m,\mu}\int \frac{dE}{\hbar\Omega}\frac{f(E)}{\pi\sqrt{\frac{E}{\hbar\Omega}}~g^{\mu}}&&\sin^2(\pi n_m^{\mu})\nonumber\\
\times&&\left[\left(n_{\overline{m}}^{\mu}-\frac{\Phi}{\Phi_0}\right)\beta^{\mu}+\gamma^{\mu}\right],\nonumber\\
\end{eqnarray}
with
\begin{eqnarray}
\gamma^{1}&=&\sin^2\frac{\theta}{2};~~~~\gamma^2=-\cos^2\frac{\theta}{2},\nonumber\\
\beta^1&=&\cos\theta;~~~~~~\beta^2=1.\nonumber
\end{eqnarray}

\begin{figure}[ht]
\begin{tabular}{cc}
\includegraphics[scale=0.85]{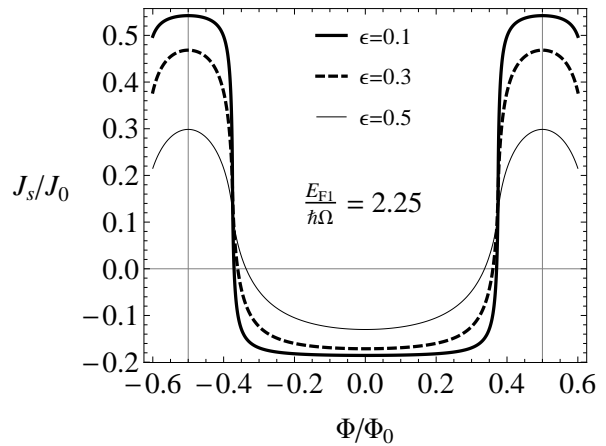} \\
\\
\includegraphics[scale=0.85]{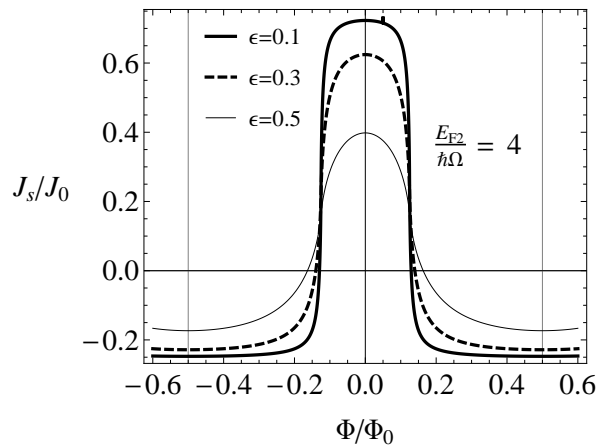}
\end{tabular}
\caption{\label{fig:SCR} Spin persistent current as a function of the magnetic flux for three values of $\varepsilon$ and $T=0$. The number of electrons is $6$ (top) and $8$ (bottom). The RSO is ${{\omega }_{SO}}/\Omega =0.75$. The current is given in untis of $J_0=\hbar\Omega$. }
\end{figure}

\begin{figure}
\includegraphics[scale=0.5]{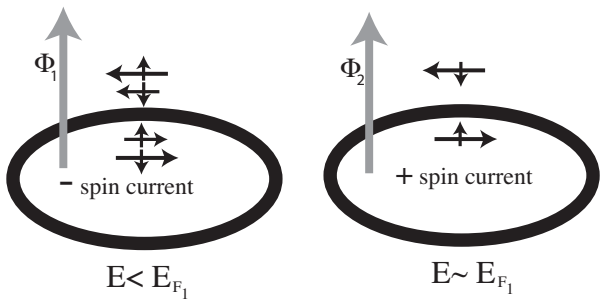} 
\caption{\label{fig10} The figure depicts, qualitatively, the contributions to the spin current as the flux changes until the Fermi level is reached. On the left, the currents in each direction are highly compensated in spin (each current direction contains both spin directions). On the right, the flux is such that the energy is close to the Fermi level $E_{F_1}$, and the spin current is large and switches direction. }
\end{figure}

Figure~\ref{fig:SCR} depicts the spin persistent current as a function of the magnetic flux for the two Fermi levels considered in Fig.\ref{fig:DeS}. Spin currents are only possible in the presence of SO coupling, since spin degeneracy matches up identical contributions in charge current from opposite spins (see Fig.\ref{fig:DeS} top panel). In the presence of the SO coupling there is a breaking of spin degeneracy with preservation of the time reversal symmetry, the necessary ingredients for their presence. As for charge currents, spin currents from deep levels in the Fermi sea, also tend to cancel but in a more complicated fashion.  Figure~\ref{fig10} shows the combinations of charge currents with their corresponding spin orientations for the first Fermi level scenario: Deep in the Fermi sea charge currents are also paired up in spin but with small differences in electron velocities due to broken degeneracy. So we can see a small spin current accrued coming from these levels. As one goes higher in magnetic field the positive current levels slow down, making less of a contribution, while the level with negative charge currents speed up, making the bulk of the current. The dispersion being quadratic makes for precise compensation, so that the full spin current is constant.

When the flux is large enough for the levels to cross the Fermi level, there is an abrupt disappearance of the negative spin up current and
a new contribution from a positive spin up charge current, as shown in Fig.\ref{fig10} right panel. These two contributions make for a pure spin current, more than three times the magnitude of the previous regime, very close to the Fermi level $E_{F_1}$. 
The range of fluxes in which this happens is as wide as it takes for the second level to emerge from the Fermi sea i.e. $\Delta(\Phi/\Phi_0)=\sqrt{1+(\omega_{SO}/\Omega)^2}-1$, at which point we start with the scenario on the left panel and repeat the whole periodic oscillation. 
\begin{figure}[H]
\begin{tabular}{cc}
\includegraphics[scale=0.85]{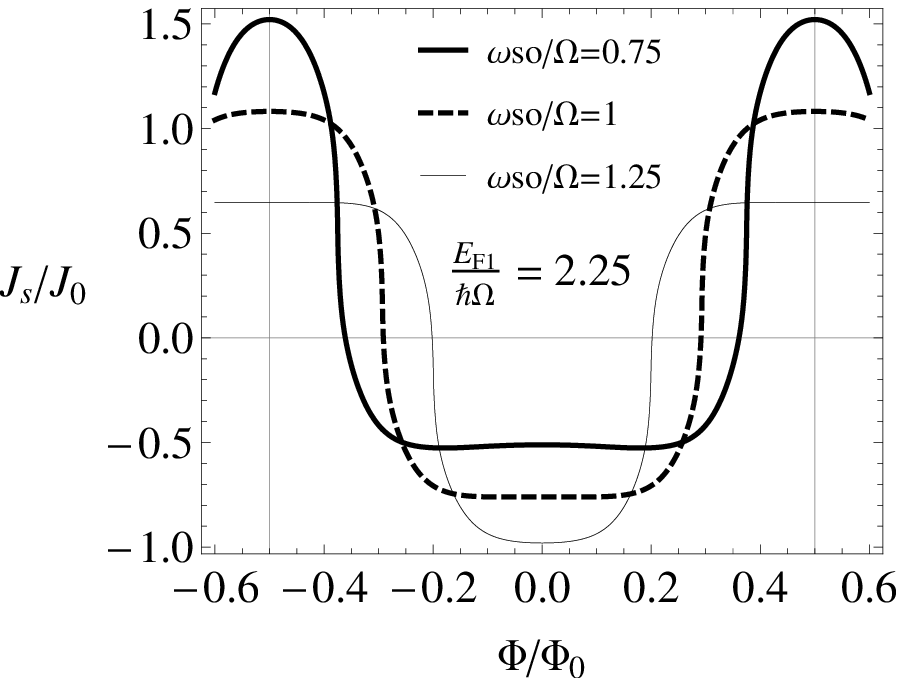} \\
\\
\includegraphics[scale=0.85]{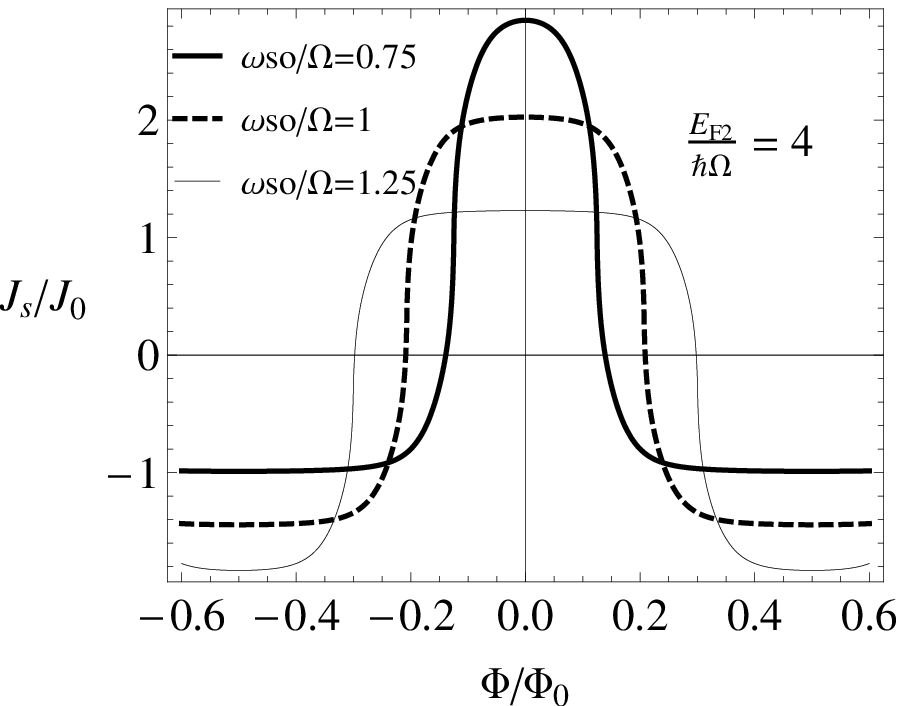}
\end{tabular}
\caption{\label{fig:SCTso} Spin persistent current as a function of the magnetic flux for three values of $\omega_{so}/\Omega$ with finite temperature, $T/T_0=0.1$ and $\epsilon=0.1$. The number of electrons is $6$ (top) and $8$ (bottom).}
\end{figure}
For the second Fermi level the scenario is identical but it occurs for small fluxes in the center of the spectrum (Fig.\ref{fig:SCR} bottom panel). Figure~\ref{fig:SCTso} shows how the spin currents, coming from different parts of the spectrum explored by the magnetic flux, can be tuned by the spin-orbit interaction at fixed coupling to the reservoir. One can see how positive and negative spin currents can be enhanced and change the range of fluxes for which they arise.

It is interesting to note that the smaller spin current coming from levels deeper in the Fermi sea is more robust to decoherence (affected less by coupling to the reservoir) than the contributions coming from close to the Fermi level, resembling thermal effects previously discussed. On the other hand, as discussed for the charge currents, the B\"uttiker model is unable to completely degrade spin currents.

\section{Summary and conclusions}

The robustness of devices involving spin manipulation through the SO coupling against decoherence and thermal effects 
is crucial for their feasibility, since these effects are unavoidable in practical applications. The latter, with the current lithographic
techniques, always involve voltage gates, contacts with external reservoirs and temperature points of operation which
should not compromise the spin sensitive physics of the device. With this concern in mind we have solved for a generalization 
of the B\"uttiker voltage probe model in SO active rings threaded by a magnetic flux. The 
procedure involves the determination of a complete set of basis functions for the uncoupled, phase coherent, problem and
then relaxing the quantization conditions on the closing of the wave functions when the scattering conditions are met at the
reservoir junction. The coupling to the reservoir is spin insensitive and the  thermal effects only determine the electron 
filling of the ring and did not account for additional broadening of the energy levels. 

Complete analytical expressions for the density of states are obtained as a function of energy, magnetic flux and
SO coupling. As expected, the isolated ring levels broaden, and they do so in an energy dependent fashion as the
reservoir couples optimally at its own Fermi energy. We note that broadening effects are only Lorentzian for weak
coupling to the reservoir, thus our results  contemplate strong reservoir coupling regime.

The equilibrium charge currents and spin currents where computed as sensitive probes for the action of both reservoir coupling
and thermal effects. The linear response formula to derive such currents is not directly useful in this case since the energy
levels broaden into a continuum, so the quantum mechanical definitions were used with the full knowledge of the wave functions
derived from the analytical procedure. Note that the full knowledge of the wavefunction implies that the model
reservoir only dephases, but there is no loss of information that would entail a density matrix description.

We computed the equilibrium charge and spin currents in the SO active ring coupled to the reservoir and assessed their coupling dependence to the electron reservoir and the effect of thermal occupation. Two representative Fermi level scenarios where considered, that involved where the spin split 
structure of the spectrum is critical i.e. close to multiples of $\Phi_0/2$. At those points the sawtooth oscillating equilibrium current
can be best modulated by the SO coupling strength. Experimentally feasible values for the SO strength were used in the computations.  

While the coupling to the reservoir uniformly degraded the coherent currents, the thermal effects revealed the interesting feature that there exist certain flux ranges that are protected by a dispersion dependent gap to the Fermi energy. This gap can be tailored by fixing the Fermi level and or the field flux. The magnitude of these protected currents is spectrum dependent but promise tailoring by considering more detailed models accounting for
ring thickness and edge effects\cite{Akhmerov}. Equilibrium spin currents are obtained in steplike ranges in flux only for Rashba spin-orbit active material. The currents steps are also rounded by coupling to the reservoir and temperature effects. Nevertheless, as these currents are built from charge currents distinguished in spin, so they are endowed with the same protective gaps. Therefore there is a range of fluxes where spin currents are thermally protected. We expect the phenomena borne out from our model to be readily checked and exploited experimentally in recent
techniques such as cantilever  torsional magnetometry\cite{Shanks2}.

\section*{Acknowledgment} NB thanks the Coll\`ege Doctoral Franco-Allemand 02-07 ``Physics of Complex Systems'' for financial support, 
EM and BB are respectively grateful to the University of Lorraine and to IVIC for invitations. They also thank the CNRS for support through the ``PICS'' programme {\it Spin transport and spin manipulations in condensed matter: polarization, spin currents and entanglement}. EM acknowledges support from Fundaci\'on POLAR.


\begin{thebibliography}{50}
\bibitem{Winkler} R. Winkler, {\it Spin-Orbit Coupling Effects in Two Dimensional Electron and
Hole Systems} (Springer-Verlag Berlin, Heidelberg, New York, 2003).
\bibitem{SpinControlElectrical} J. R. Petta et al, Science {\bf 309}, 2180 (2005).
\bibitem{SpinDecoherence} A. V. Khaetskii, D. Loss, L. Glazman, Phys. Rev. Lett. {\bf 88}, 186802 ( 2002);
\bibitem{Nitta} J. Nitta, F. E. Meijer, an d H. Takayanagi, Appl. Phys. Lett. {\bf 75}, 695  (1999); F. E. Meijer, A. F. Morpurgo, and T. M. Klapwijk, Phys. Rev. B {\bf 66}, 033107 (2002).
\bibitem{Kane} M. Z. Hasan, and C. L. Kane Rev. Mod. Phys. {\bf 82} 3045 (2010).
\bibitem{Governale} S. Janine, M. Governale, and U. ZŸlicke, Physs Rev. B {\bf 68} 65341 (2003).
\bibitem{Frustaglia} D. Frustaglia, and K. Richter, Phys. Rev. B {\bf 69}, 235310 (2004).
\bibitem{Ionicio} R. Ionicioiu and I. DÕAmico, Phys. Rev. B 67, 041307(R) (2003).
\bibitem{Hatano}N. Hatano, R. Shirasaki, and H. Nakamura, Phys. Rev. A {\bf 75}, 032107 (2007); B. Santos, E. Medina, A. Lopez, and B. Berche, J. Appl. Phys. {\bf 110}, 114523 (2011);  
\bibitem{Duan} Duan-Yang Liu, and Jian-Bai Xia, Journal of Applied Physics {\bf 115}, 044313 (2014).
\bibitem{Gudmundsson} M. Nita, D. C. Marinescu, A. Manolescu, V. Gudmundsson, Phys. Rev. B {\bf 83}, 155427 (2011); M. Nita, D. C. Marinescu, A. Manolescu, B. Ostahie, and V. Gudmundsson, Physica E, {\bf 46} 12 (2012).
\bibitem{Dedkov} Y. S. Dedkov, M. Fonin, U. Rudiger, and C. Laubschat, Phys. Rev. Lett. {\bf 100}, 107602 (2008); M. Zarea, and N. Sandler, Phys. Rev. B {\bf 79}, 165442 (2009).
\bibitem{Marchenko} D. Marchenko et al, Nature Comm., {\bf 3}, 1232 (2012).
\bibitem{Buttiker} M.~B\"uttiker, Phys. Rev. B.  {\bf 32}, 1846 (1985); M. B\"uttiker, IBM J. Res. Dev. {\bf 32}, 63, (1988).
\bibitem{Pareek} T. P. Pareek, S. K. Joshi, and A. M. Jayannavar, Phys. Rev. B 57, 8809 (1998). 
\bibitem{Tsymbal} E. Y. Tsymbal, V. M. Burlakov, and I. I. Oleinik, Phys. Rev. B 66, 073201 (2002). 
\bibitem{LiYan} X.-Q. Li and Y. Yan, Phys. Rev. B 65, 155326 (2002). 
\bibitem{Golizadeh} R. Golizadeh-Mojarad and S. Datta, Phys. Rev. B 75, 081301 (2007). 
\bibitem{Sheng} J. S. Sheng, and K. Chang, Phys. Rev. B {\bf 74}, 235315 (2006).
\bibitem{Margulis} V. A. Margulis, and V. A. Mironov, Physica E {\bf 43} 905 (2011).
\bibitem{RingTB} S. K. Maiti, M. Dey, S. Sil, A. Chakravarti, and S. N. Karmakar, Europhys. Lett. {\bf 95}, 57008 (2011).
\bibitem{Berch} B.~Berche, C.~Chatelain, and E.~Medina, Eur. J. Phys.  {\bf 31}, 1267 (2010). 
\bibitem{Bolivar}  N. Bolivar, E. Medina, and B. Berche, Phys. Rev. B, {\bf 89}, 125413 (2014).
\bibitem{reviewonprobes} H. F\"orster, P. Samuelsson, S. Pilgram, and M. B\"uttiker, Phys. Rev. B {\bf 75}, 035340 (2007).
\bibitem{Buttiker2} M.~B\"{u}ttiker, Y.~Imry and M. Ya. Azbel. Phys. Rev. A.  {\bf 30}, 1982 (1984). 
\bibitem{Pastawski} H. M. Pastawski, and E. Medina, Rev. Mex. Fis. {\bf 47} S1, 1 (2001). arXiv preprint cond-mat/0103219.
\bibitem{Datta} S. Datta and B. Das, Appl. Phys. Lett. {\bf 56}, 665 (1990).
\bibitem{Sangchul} S. Oh, and C. M.~Ryu, Phys. Rev. B.  {\bf 51}, 13441 (1995).
\bibitem{Ekenberg} U. Ekenberg, and D. M. Gvozdic. Phys. Rev. B.  {\bf 78}, 205317 (2008).
\bibitem{Daniel} D. Gos‡lbez-Mart'nez, et al. Sol. State Comm. {\bf 152} 1469 (2012).
\bibitem{Shanks} A. C. Bleszynski-Jayich, W. E. Shanks, and J. G. E. Harris, Appl. Phys.
Lett., {\bf 92}, 013123 (2008).
\bibitem{Molnar} B.~Moln\'ar, F. M. Peeters and P. Vasilopoulos. Phys. Rev. B.  {\bf 69}, 155335 (2004).
\bibitem{Akhmerov} J. A. M. van Ostaay, A. R. Akhmerov, C. W. J. Beenakker, and M. Wimmer, Phys. Rev. B 84, 195434
(2011).
\bibitem{Shanks2} A. C. Bleszynski-Jayich, W. E. Shanks, B. R. Ilic, and J. G. E. Harris, J. Vac. Science \& Technology B, {\bf 26},
1412, (2008).









               
\end{thebibliography}
\end{document}